\documentclass[preprint,showpacs,preprintnumbers,amsmath,amssymb]{revtex4}

\usepackage{graphicx}
\usepackage{bm}

\begin{document}

 \title{Quantum corral wave function engineering}

 \author{Alfredo A. Correa} \affiliation{ Department of Physics,
   University of California at Berkeley, Berkeley, California 94720,
   USA \\ Lawrence Livermore National Laboratory,
   Livermore, California 94550, USA } 
 \email{correaa@socrates.berkeley.edu}

 \author{Fernando A.  Reboredo} 
 \affiliation{ Lawrence Livermore National Laboratory,
   Livermore, California 94550, USA } \email{reboredo1@llnl.gov}
 \author{C. A. Balseiro} 
 \affiliation{ Centro At\'omico Bariloche Instituto Balseiro,
   S.C. de Bariloche, R\'{\i}o Negro 8400, Argentina }

\date{\today}

 \begin{abstract}
   We present a theoretical method for the design and optimization of
   quantum corrals with specific electronic properties.  Taking
   advantage that spins are subject to a RKKY interaction that is
   directly controlled by the scattering of the quantum corral, we
   design corral structures that reproduce spin Hamiltonians with
   coupling constants determined a priori.  We solve exactly the
   two-dimensional electron gas scattering problem for each corral
   configuration within the effective mass approximation and s-wave
   scattering using a Green function method.  Subsequently, the
   geometry of the quantum corral is optimized with an algorithm that
   combines simulated annealing and genetic approaches. We demonstrate
   that it is possible to automatically design quantum corrals with
   complicated target electronic properties, such as multiple mirages
   with predefined relative intensities at specific locations. In
   addition we design structures that are particularly sensitive to the
   phase shift of impurities at certain positions allowing the
   measurement of the value of this parameter on the copper surface.
 \end{abstract}

                              

\maketitle

\section{Introduction}

The fabrication of solid state quantum computing devices involves both the
ability to manipulate mater at the nanoscale \cite{kane:98}
 and to design structures
that interact according to prescribed quantum computing
Hamiltonians\cite{divincenzo:339}.  Tuning and controlling quantum
interactions are basic problems faced by the quantum computing community.
This requires engineering the structure of electronic wave
functions in different environments and conditions.  The manipulation
of individual atoms with STM has made possible the construction of
arbitrary quantum structures on top of surfaces. In particular, after
the seminal work of Crommie et al.\cite{crommie:93}, it has been
possible to build \emph{quantum corrals} of arbitrary shape, placing
atoms one at a time.  Quantum corrals are a collection of atoms arranged
in a controlled manner on top of a metallic surface.  These novel
structures generate quantum confinement of the surface conduction
electron wave functions leading to striking phenomena such as
resonant electronic states and the formation of \emph{ quantum
  mirages} \cite{manoharan:00}.  Generally speaking, quantum mirages
are the projection of a perturbation on a point into another distant
point of the surface. Manoharan et al.\cite {manoharan:00} were able
to partially project the Kondo \cite{hewson} electronic structure up
to $112\mathrm{\AA }$ away from the actual Kondo impurity. In these
experiments, the formation of a Kondo mirage relies on the focusing
properties of elliptical corrals.  These findings triggered a
collection of theoretical papers modeling\cite
{agam:484,aligia:121102,hallberg:066802} and extending the concept of
quantum mirages\cite{correa:899,chiappe:075421}.

This new phenomenon has been proposed as a tool for remote probing
\cite {manoharan:00} and as a way to enhance
\cite{hallberg:066802,correa:899} the interaction between localized
magnetic impurities on a surface.  Unfortunately, the design of quantum
corrals with certain desired properties is still a matter of trial and
error\cite{manoharan:00}. We think that both technological
applications and novel physical phenomena could emerge if one could
tailor the electronic wave functions that are responsible for the
quantum mirage formation. For that purpose we believe that new
systematic methods to design quantum corrals are required.

In previous works it was shown that the magnetic interaction between
magnetic impurities can be strongly enhanced due to the electronic
confinement produced by quantum
corrals\cite{hallberg:066802,correa:899}. Our present objective is to
generate the optimal geometrical arrangement of quantum corral
atoms that achieves an arbitrary predefined quantum mechanical
interaction.  This type of problem is usually denoted as \emph{inverse
  problem}. In contrast to the \emph{direct problem}, where a physical
property is the result of the geometry of the material, in the
\emph{inverse problem} the optimized geometry is a result of the
physical property targeted. Inverse problems have been addressed in a
variety of systems ranging from band gaps of solids
\cite{franceschetti:60}, design of antennas\cite{lohn:01}, or
intermolecular potential fitting\cite{globus:03}.  In order to solve
the inverse problem one needs two ingredients: i)~a fast method to
solve the direct problem (i.e.  electron gas response for a given
corral configuration) ii)~an efficient algorithm to select new
configurations and optimize the structure to achieve the target
properties.

In this paper we address the `inverse quantum corral problem'. We
concentrate in the search of structures that could reproduce
predefined magnetic Hamiltonians with desired coupling constants by
means of an enhanced RKKY interaction. We demonstrate the
possibility to physically construct several Hamiltonian examples.

Our theoretical approach consists of i)~solving each trial corral
configuration with a Green function method\cite{hallberg:066802} and
ii)~optimizing the corral geometry with a combination of
\emph{simulated annealing} and \emph{genetic
algorithms}\cite{aarts}. We find that i)~it is possible to design
quantum corrals with multiple mirages with both predefined positions
and intensities once one has an accurate model of the scattering.
ii)~Our theory allow us to design quantum corrals that are
particularly sensitive to the values of the phase shift, the only free
parameter of the model. We show that it is possible to design
structures that can be used to measure this phase shift.

\section{Direct problem solution} 

\subsection{Solving the scattering}

We consider quantum corrals on the Cu(111) surface. Cu atoms on this
surface form a triangular lattice with spacing $a=2.55\mathrm{\AA }$.
This underlying triangular lattice defines also a triangular lattice
for the equilibrium positions of the impurities that form the corral.
The electronic structure of Cu(111) consist of a surface band with a
Fermi energy $ \epsilon _{F}$ laying $450~\mathrm{meV}$ above the
bottom of this band. We will treat the surface band as a
two-dimensional-electron gas in the effective mass approximation.
There is broad consensus
\cite{manoharan:00,agam:484,porras:01,aligia:121102,hallberg:066802,correa:899,fiete:933}
that the quantum mirages form as a result of the interplay between the
impurities at the surface and this two dimensional surface band.

Several types of impurity atoms have been used to construct quantum corrals,
examples are Fe \cite{crommie:93} and Co \cite{manoharan:00}. In all cases,
the Fermi wave length of the electrons in the surface band is much larger
than the spatial extension of the perturbation introduced by a single
impurity atom. Therefore, an s-wave scattering approximation is good enough
for the description of the electronic scattering by the impurities\cite
{fiete:933}. Accordingly, the impurities that form the corral are described
by a single parameter being the s-wave phase shift $\delta $.

For the case of $N$ scatters located at positions $\{{\bf a}_i\}$ the
retarded Green function $G^{\rm ret}$ of the system can be expressed
in the following way\cite{RT}:
\begin{eqnarray}\label{eq:G}
  G^{\rm ret}({\bf r},{\bf r'}) & = &  G^{\rm ret}_0({\bf r},{\bf r'}) + \\
                  &+&\sum_{ij} G^{\rm ret}_0({\bf r},{\bf a}_i)[1/t-{\bf
  G}'_0]^{-1}_{ij} G^{\rm ret}_0({\bf a}_j,{\bf r'}) \nonumber
\end{eqnarray}
where $[{\bf G}'_0]_{ij}=(1-\delta_{ij})\times G^{\rm ret}_0({\bf
  a}_i,{\bf a}_j)$ and $t$ is the t-matrix of the identical scatters;
in terms of the phase shift $\delta$, $t=i\frac{\hbar^2}{m^*}(e^{i 2
  \delta}-1)$, being $m^*$ the effective mass ($m^*=0.38 m_e$).  In
two dimensions and within the effective mass approximation
\cite{economou}, $G^{\rm ret}_0({\bf r},{\bf
  r'};\omega)=\frac{m^*}{2\hbar^2}({\rm Y}_0(k|{\bf r-r'}|)-i{\rm
  J}_0(k|{\bf r-r'}|))$, $\hbar k=\sqrt{2m^* \omega}$.  While this
method can be equally efficient for any phase shift model, for the
purpose of this study, we chose to set $\delta=i\infty$ (unless
otherwise specified) which has been determined to produce the best fit
of the experimental information for quantum corrals constructed with
Fe atoms on the cooper surface\cite{heller:464}. 

Equation (\ref{eq:G}) provides a straightforward way to obtain the exact Green
function 
after an $N\times N$ matrix inversion. For the typical number of
scatters that can be experimentally handled ($N\simeq 90$) this
method is fast enough to solve thousand of different atomic
configurations in seconds, each of them representing one trial step.

The Green function contains all the information about the solution of
the independent particle problem and can be used to compute, for
example, the local density of states (LDOS):
\begin{equation}
  {\mathcal D} ({\bf r};\omega)=-\frac{1}{\pi}{\rm Im} G^{\rm ret}({\bf r,
  r};\omega)
\end{equation}
or the perturbation caused by additional atoms, in particular magnetic
impurities.

Let us note that $G^{\mathrm{ret}}$ is a function of the atomic
positions $\{ \mathbf{a}_i \}$. In this work we address the question
of which is the corral (i.e. the set $\{ \mathbf{a}_i \}$) that gives
predefined couplings between magnetic impurities by means of
modulating the RKKY interaction.

\subsection{RKKY interaction in two dimensions}

The exchange interaction of a magnetic impurity at coordinate ${\bf
  R}_{1}$ with an electron gas with coupling constant $J_{1}$
is described by the Hamiltonian:
\begin{equation}
H_{\mathrm{imp}}=-J_{1}\mathbf{S}_{\mathrm{1}}\cdot \psi ^{\dagger }(\mathbf{R}_{
\mathrm{1}}){\bm\sigma }\psi (\mathbf{R}_{\mathrm{1}})
\end{equation}
where $\mathbf{S}_{\mathrm{1}}$ describes the impurity spin, ${\bm
  \sigma }=(\sigma _{x},\sigma _{y},\sigma _{z})$ are the Pauli
matrices and $ \psi ^{\dagger }(\mathbf{r})=(\psi _{\uparrow
}^{\dagger }(\mathbf{r}),\psi _{\downarrow }^{\dagger }(\mathbf{r}))$
describes the electron field.

If a second magnetic impurity with spin $\mathbf{S}_{\mathrm{2}}$ is
placed at $\mathbf{R}_{\mathrm{2}}$, the effective impurity-impurity
coupling mediated by the electron gas can be written as:

\begin{equation}
H_{12}=-\mathcal{J}_{12}\mathbf{S}_{1}\cdot \mathbf{S}_{2}  \label{eq:Hij}
\end{equation}
with 

\begin{equation}
\mathcal{J}_{12}=\!\frac{4J_{1}J_{2}}{\pi }\int d\omega f(\omega )\mathrm{Im}
[G_{\downarrow }^{\mathrm{ret}}(\mathbf{R}_{\mathrm{1}},\mathbf{R}_{\mathrm{2
}};\omega )G_{\uparrow }^{\mathrm{ret}}(\mathbf{R}_{\mathrm{2}},\mathbf{R}_{
\mathrm{1}};\omega )]\,.  \label{jrkky}
\end{equation}
where $J_{2}$ is the coupling of the second impurity with the electron
gas, $ \ f(\omega )$ is the Fermi function and $G_{\sigma
}^{\mathrm{ret}}(\mathbf{r },\mathbf{r^{\prime };}\omega )$ is the
retarded Green function for electrons with spin $\sigma $.

In general, to lowest order, the magnetic behavior of a collection of impurities at
coordinates $\{ {\bf R}_i \}$ is given by the following
Hamiltonian: 
\begin{equation}
  H=\sum_{\langle ij \rangle} -\mathcal{J}_{ij} {\bf S}_i \cdot {\bf S}_j
\end{equation}
where the summation is done over all impurity pairs. The exchange
parameters $ \mathcal{J}_{\mathrm{ij}}$ can be written as
$\mathcal{J}_{ij}=J_i J_j C({\bf R}_i,{\bf R}_j)$
with the correlation function
$C({\bf R}_i,{\bf R}_j)$ given by

\begin{eqnarray}
C({\bf R}_i,{\bf R}_j)\!\!\! &\!\!=\!\!\!&\frac{4}{\pi }\!\!\!\int \!\!
d\omega f(\omega )\mathrm{Im}[G_{\downarrow }^{\mathrm{ret}}({\bf R}_i,{\bf R}_j;\omega )G_{\uparrow }^{\mathrm{ret}}(
\mathbf{R}_j,\mathbf{R}_i;\omega )]  \nonumber \\
&=&-\frac{4}{\pi }\int_{-\infty }^{\epsilon _{F}}\mathrm{d}\omega \mathrm{Im}
\left[ G^{\mathrm{ret}}(\mathbf{R}_i,\mathbf{R}_j;\omega )^{2}\right]   \label{eq:c}
\end{eqnarray}
The second line is obtained after assuming zero temperature, using the
relations $G_{\downarrow
}^{\mathrm{ret}}(\mathbf{R}_{\mathrm{1}},\mathbf{R}
_{\mathrm{2}};\omega )=(G_{\downarrow
}^{\mathrm{adv}}(\mathbf{R}_{\mathrm{2}
},\mathbf{R}_{\mathrm{1}};\omega ))^{\ast }$ and taking spin
independent Green functions, this is exact in the absence of
spin-orbit coupling and external\ magnetic fields. From here on we
drop the spin index in the Green functions. The major contribution to
the integral of equation (\ref{eq:c}) comes from states near the Fermi
energy\cite{kittel}, i.e. from $\mathrm{Im} \left[
  G^{\mathrm{ret}}(\mathbf{R}_i,\mathbf{R}_j;\epsilon _{F})^{2}\right]
$ \cite{kittel,correa:899}. For the sake of avoiding the time
consuming integration of Eq. (\ref{eq:c}) in the numerical method, we
use the dominant contribution from the Fermi level:
\begin{equation}
C_{\epsilon _{F}}(\mathbf{R}_i,\mathbf{R}_j)=-\frac{4}{
\pi }\mathrm{Im}\left[ G^{\mathrm{ret}}(\mathbf{R}_i,\mathbf{R}_j;\epsilon _{F})^{2}\right] 
\end{equation}

In the free surface, the RKKY interaction is oscillatory with a power
law decay $1/\left|
  \mathbf{R}_i-\mathbf{R}_j\right| ^{2}$ . In the
presence of an appropriate surrounding corral this interaction can be
enhanced and focused\cite{correa:899}. In this work we exploit the
fact that the correlation
$C_{\epsilon_F}(\mathbf{R}_i,\mathbf{R}_j)$ can be controlled
by changing the shape of the quantum corral, i.e. the RKKY interaction
can be controlled by the quantum corral design.

Also, this correlation function (8) can be shown to be proportional to
the perturbation in the $LDOS$ at point $\mathbf{R}_{\mathrm{2}}$
produced by a non magnetic impurity located at
$\mathbf{R}_{\mathrm{1}}$ [$\delta {\mathcal D}({\bf R}_2,{\bf R}_1)$].
Therefore, to linear order, the search of magnetic mirages in the spin
density is the same than the search of mirages in 
${\mathcal D} (\mathbf{R}_{\mathrm{2}},\mathbf{R}_{\mathrm{1}})$; extending the
scope of this work to the case where the perturbation is non magnetic but
still localized in space.

\section{The inverse problem}

Suppose that $N$ impurities can fit in a lattice with spacing $a$, and
we restrict them to a square with side $l$. The total number of
possible configurations (including open geometries) is the
combinatorial $n=(l/a)^2!/((l/a)^2-N)!N!$.  For a typical system of
$N=30$, $a=2.5{\text \AA}$ and $l=100 {\text \AA}$, $n\simeq 10^{64}$.
Therefore, it is impossible to study all these configurations within
reasonable times.  Moreover, in the case of quantum corrals, most
measurable quantities are expected to be non-smooth-multiple-valley
functions of the $2N$ coordinates of the impurities due to the
appearance and disappearance of resonant states for different
configurations of the quantum corral (usually close-shaped corrals
produce sharp resonances at discrete energies\cite{crommie:93}).
Therefore with a simple relaxation scheme, the system may be trapped in local
minima of the multiple-valley function. Accordingly, more sophisticated non
deterministic methods are required.

In spite of this large number of configurations, for most practical
applications it is possible to avoid covering all the configuration space.
At the same time, it is possible to find good enough solutions than can be
improved systematically. 

\subsection{Simulated annealing}

The simulated annealing algorithm \cite{aarts,press} is a quite general
method to obtain minima of complicated functions of a large number of
variables. In order to use a simulated annealing method we must
define first a {cost function} $E(\{\mathbf{a}_{i}\})$. The cost
function is simply a mathematical expression that quantifies the
desired physical properties of the system; being minimum when the
target is reached and maximum if undesired properties are present.
Accordingly, the cost functions used in this work depend on one or
more correlation functions. Several examples will be discussed in the
applications below.

Once the cost function $E(\{\mathbf{a}_{i}\})$ is defined and a random
initial configuration is chosen, the simulated annealing algorithm
consists\cite {metropolis:53} in generating \emph{random trial steps}
and accepting them with a probability chosen following the Metropolis
algorithm:
\begin{equation}
P(\Delta E)=\left\{ {\begin{array}{*{20}c} 1 & {\text{if}\,\Delta E
\leqslant 0} \\ {\exp ( - \Delta E/T)} & {\text{if}\,\Delta E > 0} \\
\end{array}}\right. 
\end{equation}
where $\Delta E$ is the change on the cost function in the trial step
and $T$ is a \emph{fictitious temperature}.  At each trial step the
coordinates of the atoms of the corral are allowed to move in a
triangular lattice representing the
Cu(111) surface.  In the annealing algorithm this \emph{Metropolis
  dynamics} is repeated while $T$ is gradually reduced\cite{press}.
This process allows the system to sample the space of variables,
staying longer times near better global minimum as $T$ diminishes. At
$T=0$, only favorable ($\Delta E<0$) steps are allowed to take place
but at finite $T$ the system has the opportunity to escape from local
minima. Eventually, if the fictitious temperature decreases slowly
enough, the result will be a global or at least a good minimum. A
particular property of the present system is, however, that for some
parameters, as the quality of the minimum increases, the amplitude of
the barriers grows.  Therefore, rather than reducing the temperature,
we smoothly adjust it continuously so as to maintain a target number
of accepted changes.  This implies that, in our case, as the cost
function improves, the temperature rises. Finally, instead of reducing
the fictitious temperature, as in the general simulated annealing
case, we slowly reduce the target of accepted changes\cite{aarts}.

The rate of this reduction and how to define a cost function for a
particular problem is in general a matter of trial and error and is not
unique. Several examples for different situations will be given.

\subsection{Genetic algorithm}

Due to the presence of resonances, the performance of the
simulated annealing decreases as better solutions are found or as a
more complicated cost functions are introduced. Since we are
actually looking for a big resonance, we have to consider a
complementary method to optimize the configurations further. In order
to further improve the optimization we combined simulated annealing
with another Monte Carlo technique which is sometimes more robust,
a {\em genetic algorithm}\cite{aarts}. The genetic algorithm is a
learning model which derives from an analogy with \emph{evolution}
process in nature.

In our case, for the quantum corral problem, we have designed a
genetic algorithm that starts with a \emph{population} of different
corral configurations, termed \emph{individuals}.  In order to use
genetic algorithms, one must define also a cost function (that we
choose to be the same one used in the simulated
annealing).  The proposed cost function is evaluated for each
individual of the population.  The evaluation of each individual
determines its probability to survive to the next \emph{generation}.
Statistically only the fittest (lowest cost function) are chosen to
survive, those which are eliminated are replaced by the {\em
  offspring} resulting from mating survivors. In our case, we
constructed the configuration of the corrals of the offspring as a
random combination of the configurations of the corrals of the
surviving parents. Accordingly, the offprings keep geometrical
similarities with their parents.  In the general case of a genetic
algorithm, a random mutation\cite{aarts} is applied to the offspring
in order to maintain diversity in the population. In our case, we
replace this random mutation step by a simulated annealing step for
each individual described above. We record the configuration of each
individual that reaches the lowest cost during the simulated annealing
step and use it for the next iteration.  This combined process allows
us to move continuously from a pure simulated annealing to a pure
genetic algorithm depending on the number of individuals in the
genetic algorithm and temperature and steps in the simulated
annealing. Once a new population of quantum corrals is generated, the
process starts again retaining the fittest and eliminating the rest.
This combined procedure was repeated until a good solution was found.

In Table \ref{tb:parameters} we provide the parameters used to obtain
the results shown in the Figures 1 to 6.

\begin{table}
\centering
\begin{tabular}{|c|c|c|c|c|c|c|}
\hline
     & Number of & Targeted & Sa   &  GA   & GA         & Cost     \\
Case & impurities    & change   & steps& steps  & population & function \\
\hline\
1    & 60           & 1\% &$7\times10^6$  &  1       &      1         &          -9.7,  \\
2    & 90           & 1\%  &$5\times10^6$  &  1       &      1         &          -7.0     \\
3    & 120 & 1\%     &100\footnote{for each mutation}  &  $10^3$ & 50 &  -6.2       \\
\hline
\end{tabular}
\caption{Parameters and results for each of the cases studied}
\label{tb:parameters}
\end{table}

\section{Results}

\subsection{A single mirage case: Alternatives to the ellipse}

Guided by intuition we may think (as many of us did) that the ellipse
was the optimum shape for a corral to enhance a perturbation of an
impurity located at one focus of the ellipse into the other focus.
This intuitive though originates in an analogy with geometrical
optics\cite{agam:484}.  Unfortunately, the analogy is valid only
partially because of the following reasons: i)~for the characteristic
length scales of quantum corrals, geometrical optics is not completely
valid and non trivial effects can arise from multiple scattering and
purely quantum interference ; ii)~the impurities can not be freely
placed on the surface, there is always an underlying lattice (e.g.
triangular for Cu(111)) that restricts the possible locations of the
corral atoms. This restriction can have non negligible consequences as
pointed out in Ref. \cite{fiete:933}.  iii)~There is no theoretical
proof that the ellipse is the optimal configuration of corral atoms,
furthermore, the exact shape could depend on the particular phase
shift model chosen, i.e.  on the specific nature of the corral atoms.

Given these facts, the following question arises: is there any
better atomic configuration than the elliptical one?

To test our method in this simple case, the simulated annealing
algorithm was instructed to maximize the quantity $C_{\epsilon
  _{F}}({\bf R}_1,{\bf R}_2)$, i.e. to project a perturbation from
${\bf R}_1$ to ${\bf R}_2$ as efficiently as possible. The cost
function was chosen to be simply $E_1 =-C_{\epsilon _{F}}({\bf R}_1,{\bf
  R}_2)$.

The reflection symmetry at the $x$-axis was imposed on the corral
impurities to simplify the problem.  The distance between the spins at
${\bf R}_1$ and ${\bf R}_2$ was fixed to $140~{\rm \AA}$ to resemble
the experimental conditions.

\begin{figure}

\includegraphics[totalheight=0.8\textheight,clip=true]{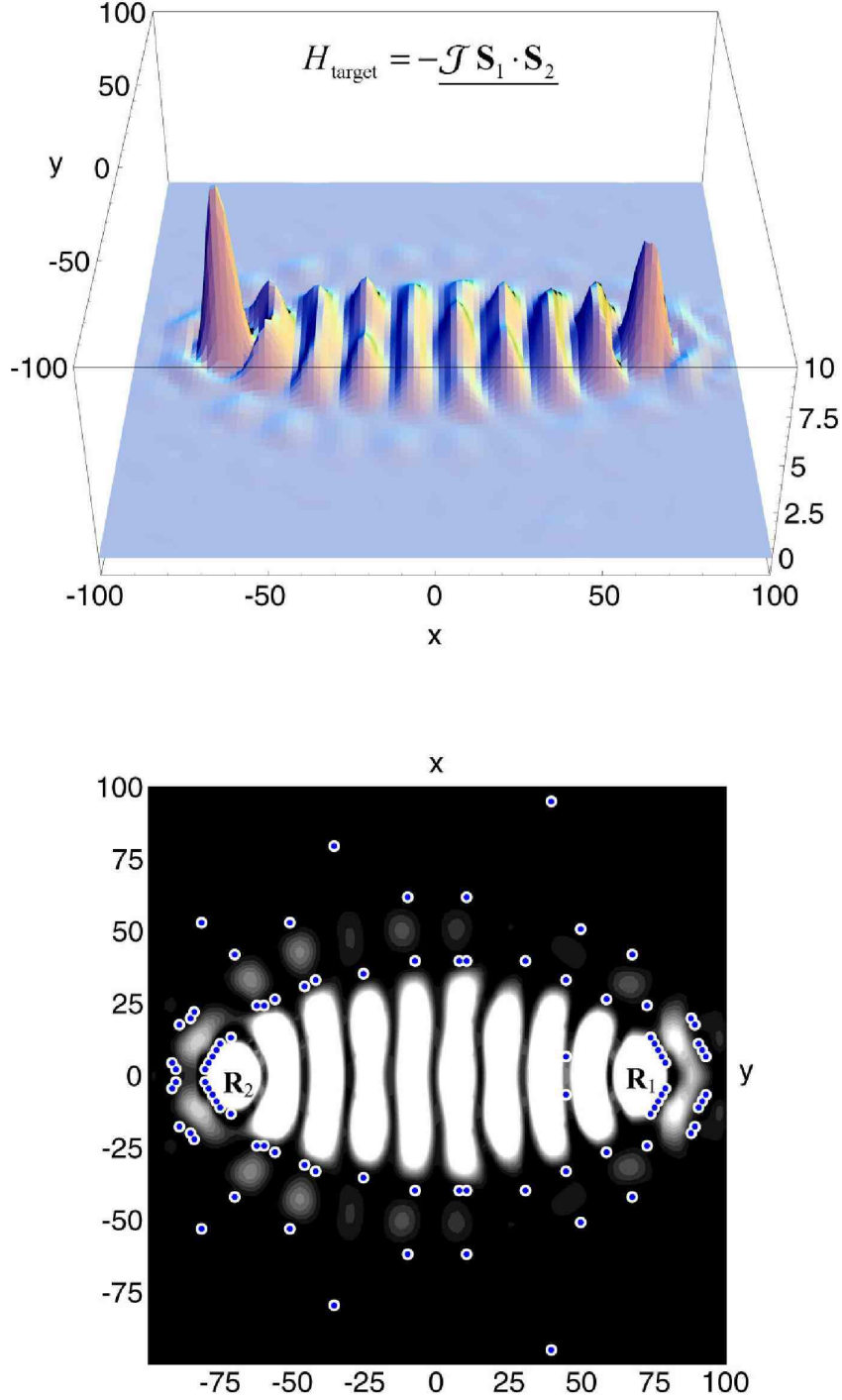}

\caption{ (Color online).
  a) Magnetic response (linear order) to a spin located at
  $\mathbf{R}_{1}$ ($x=70~\mathrm{ \AA }$, $y=0$) or quantum mirage
  formed as a result of an impurity at $\mathbf{R}_{1}$ Distances are
  reported in $\mathrm{ \AA}$. The corral is optimized to enhance the
  perturbation at point $\mathbf{R}_{2}$ ($x=-70~\AA $, $y=0$).  The
  underlined term in Hamiltonian in the inset shows which magnetic
  interaction is associated with the correlation function displayed.
  b) Structure of $C_{\epsilon_F}(\mathbf{R}_1,\mathbf{r})$ below 5\%
  of the largest peak. The design of the corral (white and blue dots)
  has been done with the Monte Carlo algorithm (see Case 1 in Table
  \ref{tb:parameters}). }
\label{fig:Two}
\end{figure}

Once the optimization procedure is completed, we plot the correlation
function of equation (7).  In Figure \ref{fig:Two} we show the
response function and the quantum corral resulting form our
optimization process [see Table \ref{tb:parameters} case 1 for
parameters].  We see in Fig.  \ref{fig:Two} that the minimum found
presents some resemblances to the ellipse but new unexpected features
also appear. First of all, the atoms of the corral tend to accumulate
around the location of the perturbation and also around the location
of the desired mirage.  Atoms on the long sides have less importance
and are placed in a less dense arrangement.  This contrasts with the
evenly spaced atoms of the early experimental setups
\cite{crommie:93,manoharan:00}.  A second shell of impurities appears
to be more efficient than a single shell to confine the electron gas
and produces better mirages.

For comparison we have studied the family of con-focal ellipses that
can be formed with the same number of equidistant impurities (being
the perturbation at one focus and the mirage at the other).  We find
that the best mirage formed by the family is a factor 3 smaller than
the one obtained with the structure shown in Fig. \ref{fig:Two}.
  
This comparison with the ellipse demonstrates the power of the
optimization technique we have developed, finding an alternative and
better configuration for the formation of single quantum mirage. In
the rest of this paper we exploit this technique further to study more
complex problems.

\subsection{Design of a quantum corral to generate chosen couplings between
  three spins}

Returning to the problem of constructing physically an arbitrary
Hamiltonian [see Eq. (6)], suppose now that we want to build a quantum
corral such that generates \emph{a target magnetic Hamiltonian} of
three impurities:
\begin{equation}
\label{eq:3spins}
H_{\rm target} = -{\cal J} {\bf S}_1 \cdot {\bf S}_2 - {\cal J} {\bf S}_1 \cdot {\bf S}_3
\end{equation}
i.e. only a selected set of pairs of the three spins are coupled. We
arbitrarily choose the coordinates of the spins $ \mathbf{R}_{1}$,
$\mathbf{R}_{2}$ and $\mathbf{R}_{3}$ to be in the vertexes of an
equilateral triangle of side $d=121~\mathrm{\AA }$.  Note that i) Eq.
(\ref{eq:3spins}) is a particular case of Eq.  (6) where
$\mathcal{J}_{23}=0$ and $\mathcal{J}_{12}=\mathcal{J
}_{23}=\mathcal{J}$; and ii) the target Hamiltonian has a different
symmetry than the desired positions of the magnetic spins. That is, we
require the magnetic couplings on two sides of the triangle but not on
the third.
In terms of the electronic spin correlation function $%
C_{\epsilon _{F}}$, this system requires $C_{\epsilon _{F}}(\mathbf{R}%
_{2},\mathbf{R}_{3})=0$ and $C_{\epsilon _{F}}(\mathbf{R}_{1},\mathbf{R}%
_{2})=C_{\epsilon _{F}}(\mathbf{R}_{1},\mathbf{R}_{3})$ to be both as
large as possible in order to enhance the interaction $\mathcal{J}$.
We propose a cost function $E_2$ that favors these two requirements that
is going to be minimized by the Monte Carlo techniques described
earlier:
\begin{eqnarray}
E_2 &=&-C_{\epsilon _{F}}(\mathbf{R}_{1},\mathbf{R}_{2})-C_{\epsilon _{F}}(%
\mathbf{R}_{1},\mathbf{R}_{3})  \nonumber  \label{eq:threespins} \\
&&+|C_{\epsilon _{F}}(\mathbf{R}_{1},\mathbf{R}_{2})-C_{\epsilon _{F}}(%
\mathbf{R}_{1},\mathbf{R}_{3})|  \nonumber \\
&&+|C_{\epsilon _{F}}(\mathbf{R}_{2},\mathbf{R}_{3})|
\end{eqnarray}

The first two terms in Eq. (\ref{eq:threespins}) ensure that the
interactions $\mathcal{J}$ are enhanced; the third term is used to
favor $\mathcal{J}_{12}$ and $ \mathcal{J}_{13}$ to be equal; the last
term penalizes the interaction $\mathcal{J}_{23}$.
If the positions of the spins $\mathbf{R}_{1}$, $\mathbf{R}_{2}$ and
$\mathbf{R}_{3}$ are selected a priori and fixed, the cost function
$E_2$ is just a function of the $2N$ coordinates $\{\mathbf{a}_{i}\}$
of the atoms forming the quantum corral. We also impose here the
reflection symmetry along the $x$ axis. The Monte Carlo algorithm
automatically minimizes this cost function $E_2$ by varying
$\{\mathbf{a}_{i}\}$. The parameters of the calculations can be taken
from Table \ref{tb:parameters} case 2.

\begin{figure}[tbp]
\includegraphics[totalheight=0.8\textheight,clip=true]{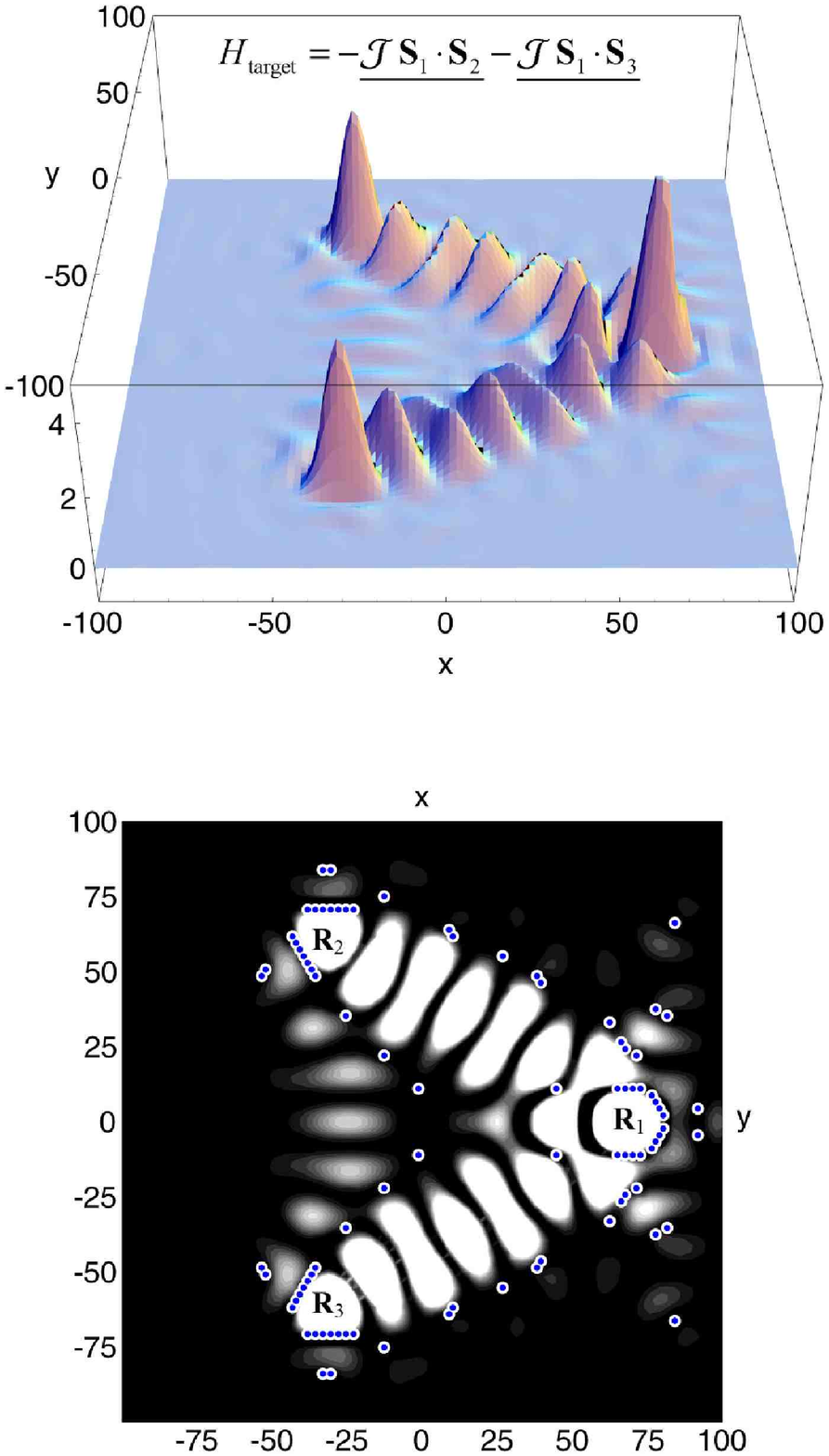}
\caption{(Color online) Quantum corral optimized to generate the spin 
Hamiltonian of Eq. (\ref{eq:3spins}).
  Response function $C_{\protect\epsilon
    _{F}}(\mathbf{R}_{1},\mathbf{r})$ due to a spin located at
  $\mathbf{R}_{1}$ ($x=70~\mathrm{\AA },y=0$). Note the magnetic
  mirages at $\mathbf{R}_{2}$ and $\mathbf{R}_{3}$. The underlined
  terms in the Hamiltonian in the inset show the magnetic interactions
  associated with the correlation function displayed.  Same
  conventions and symbols as in Fig~\ref{fig:Two}}
\label{fig:ThreeBlockR1}
\end{figure}

\begin{figure}
\includegraphics[totalheight=0.8\textheight,clip=true]{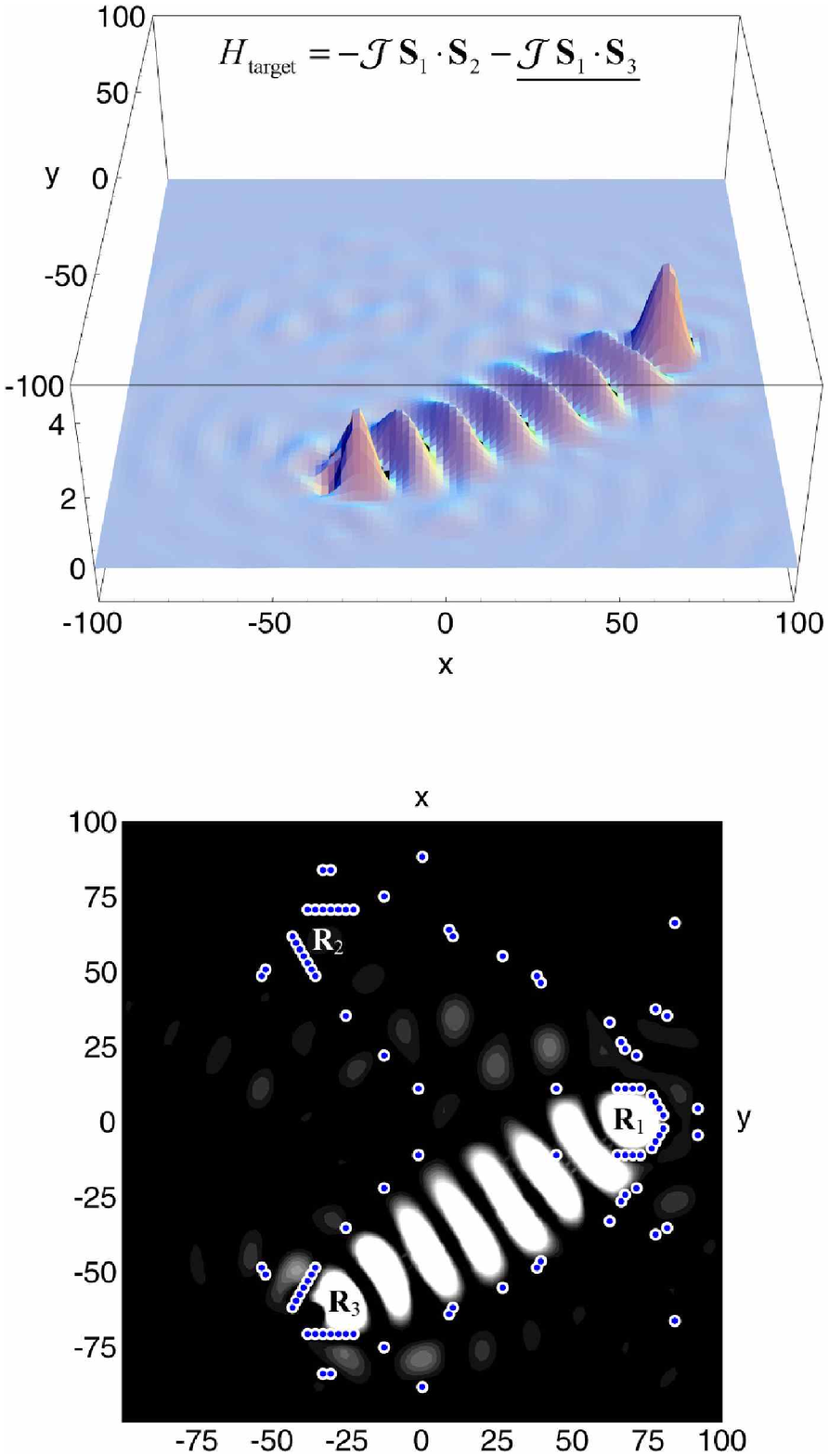}
\caption{(Color online) Quantum corral optimized to generate the spin Hamiltonian of Eq. (\ref{eq:3spins}).
  Magnetic response function $C_{\protect\epsilon
    _{F}}(\mathbf{R}_{3},\mathbf{r})$ due to a spin located at
  $\mathbf{R}_{3}$. Same conventions and symbols as in Figs.
  ~\ref{fig:Two} and ~\ref{fig:ThreeBlockR1}}
\label{fig:ThreeBlockR3}
\end{figure}

The resulting corral is shown in Fig. \ref{fig:ThreeBlockR1} and \ref
{fig:ThreeBlockR3}, the original requirements are efficiently
achieved.  In Fig.~\ref{fig:ThreeBlockR1} we see that the optimum
configuration found for the impurity atoms has dense focusing
structures near the source of the perturbation and near the mirages.
Amazingly, some atoms are automatically located in the middle of the
corral to split the standing waves towards the ``targets'' at ${\bf
  R_{2}}$ and ${\bf R_{3}}$. In Figure \ref{fig:ThreeBlockR3} we show the effect
of a magnetic impurity at $\mathbf{R}_{3}$ which produces a large
response at $\mathbf{R }_{1}$ but not at $\mathbf{R}_{2}$ [in contrast
with the case of an impurity located at $\mathbf{R}_{1}$ (compare with
Fig.  \ref{fig:ThreeBlockR1})].  Let us emphasize that the corral is
the same in Fig. \ref{fig:ThreeBlockR1} and  \ref{fig:ThreeBlockR3},
the difference is the position of the magnetic perturbation.

Comparison of Figures \ref{fig:ThreeBlockR1} and
\ref{fig:ThreeBlockR3} shows that the RKKY interaction will induce
couplings as initially designed. The resulting magnetization shows
that $\mathbf{S}_2$ and $\mathbf{S}_3$ are not
directly coupled while at the same time the interactions between $\mathbf{S}%
_1$ and $\mathbf{S}_2$ and between $\mathbf{S}_1$ and $\mathbf{S}_3$
are enhanced. We emphasize that the positions of the
spins and the Hamiltonian chosen are arbitrary. We believe that
we can find quantum corrals for a large number
of targeted Hamiltonians using the same approach.



\subsection{Four spins Hamiltonians}

In this section we show the resulting design for four spins with a target
Hamiltonian: 
\begin{equation}
\label{eq:4spins}
  H_{\rm target}=-{\cal J} {\bf S}_1 \cdot {\bf S}_3 - {\cal J} {\bf S}_2 \cdot {\bf S}_4
\end{equation}
where the positions of the spins form a
square.

In order to achieve the target Hamiltonian of Eq. (\ref{eq:4spins}) by
means of modulating the RKKY interaction, we chose the following cost
function:
\begin{eqnarray}
\label{eq:fourspins}
E_3 &=&-C_{\epsilon _{F}}(\mathbf{R}_{1},\mathbf{R}_{3})-C_{\epsilon _{F}}(%
\mathbf{R}_{2},\mathbf{R}_{4})  \nonumber \\
&&+|C_{\epsilon _{F}}(\mathbf{R}_{1},\mathbf{R}_{3})-C_{\epsilon _{F}}(%
\mathbf{R}_{2},\mathbf{R}_{4})|  \nonumber \\
&&+|C_{\epsilon _{F}}(\mathbf{R}_{1},\mathbf{R}_{2})|  \nonumber \\
&&+|C_{\epsilon _{F}}(\mathbf{R}_{2},\mathbf{R}_{3})|  \nonumber \\
&&+|C_{\epsilon _{F}}(\mathbf{R}_{3},\mathbf{R}_{4})|  \nonumber \\
&&+|C_{\epsilon _{F}}(\mathbf{R}_{4},\mathbf{R}_{1})|
\end{eqnarray}


The first two terms were introduced to increase the interaction
between spins at opposite vertexes; the role of the third term is to
favor identical coupling between the two pairs of spins (${\cal
J}_{13}={\cal J}_{24}={\cal J}$, note that any difference will
increase the cost function); the rest of the terms penalize the
coupling between consecutive spins (in order to achieve that ${\cal
J}_{12}={\cal J}_{23}={\cal J}_{34}={\cal J}_{41}=0$). For this case we
impose symmetry along the ``x'' and ``y'' axes. The method for
optimization is similar to the one used in the cases described above
but in this case the genetic algorithm plays a more important
role. The parameters of the calculations can be taken from Table
\ref{tb:parameters} case 3.

\begin{figure}
\includegraphics[totalheight=0.8\textheight,clip=true]{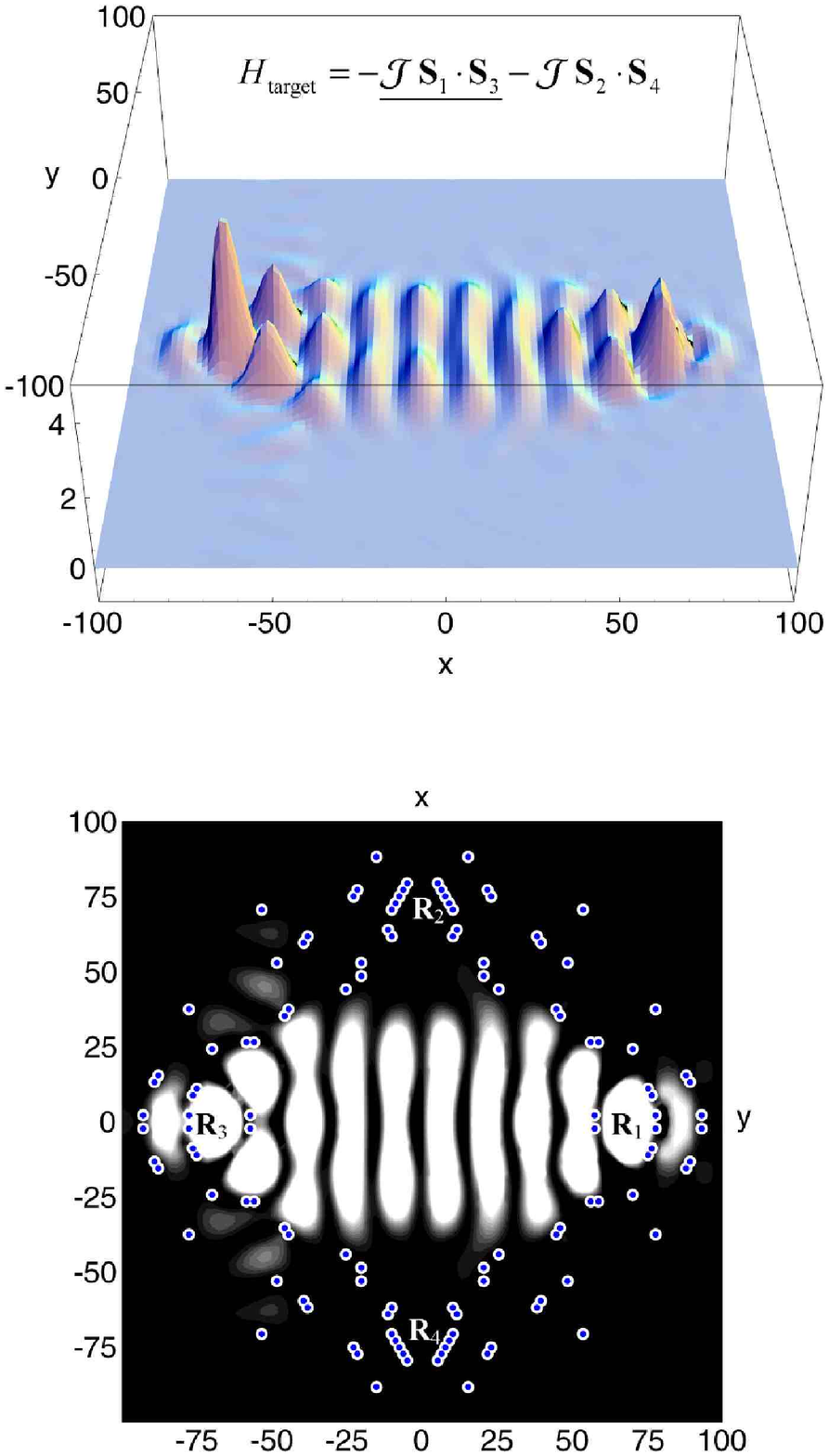}
\caption{
Quantum corral optimized to generate the spin Hamiltonian of Eq. (\ref{eq:4spins}).
Magnetic response function $C_{\protect\epsilon
  _{F}}(\mathbf{R}_{1},\mathbf{r})$ due to a spin
at $\mathbf{R}_{1}$ ($x=70~\mathrm{\AA }$,
$y=0$). Same conventions and symbols as in Fig.~\ref{fig:Two}
 Compare to Fig.
\ref{fig:FourCrossR2}.}
\label{fig:FourCrossR1}
\end{figure}

\begin{figure}
\includegraphics[totalheight=0.8\textheight,clip=true]{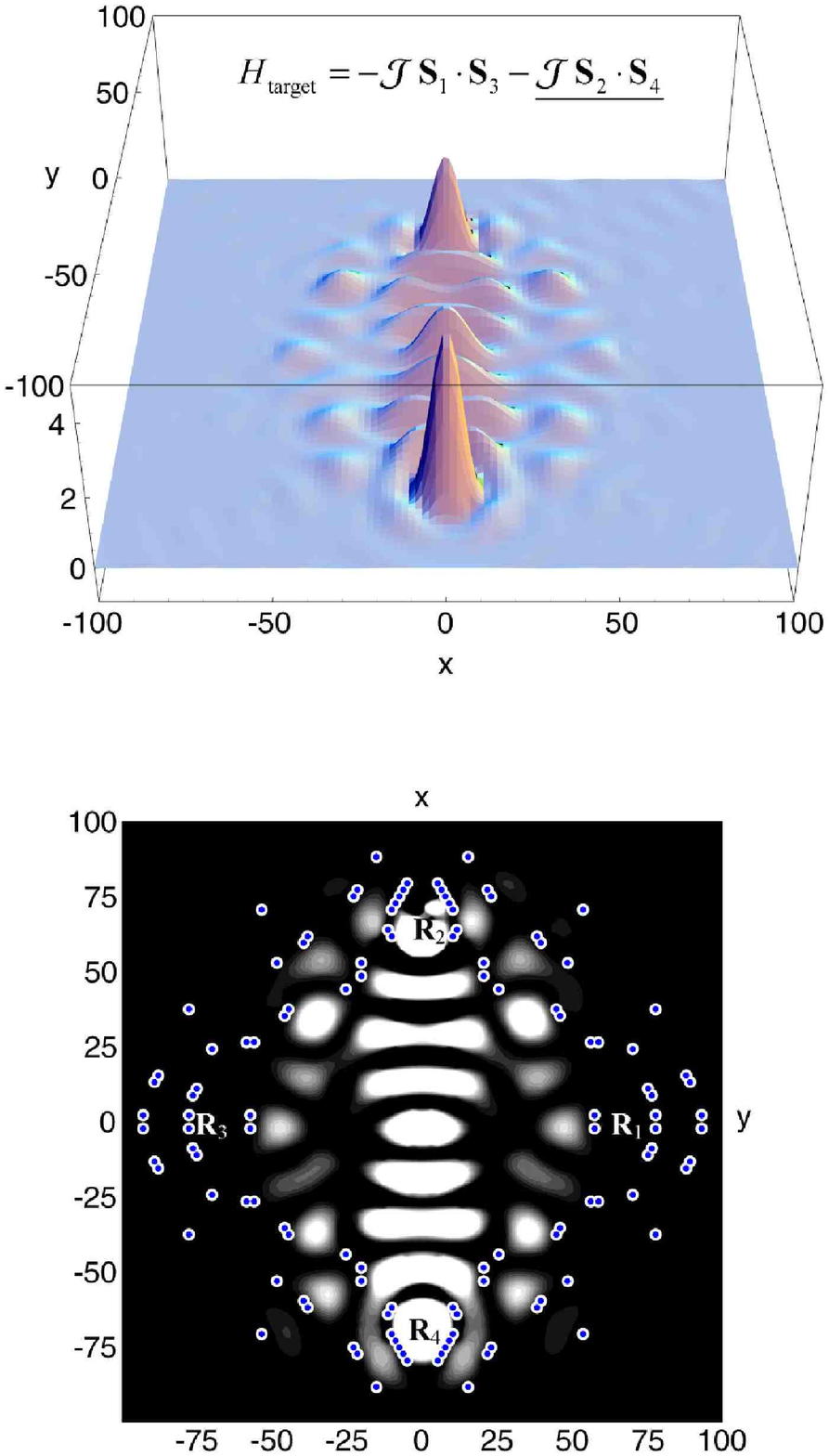}
\caption{(Color online) 
  Quantum corral optimized to generate the spin Hamiltonian of Eq.
  (\ref{eq:4spins}).  Magnetic response function $C_{\protect\epsilon
    _{F}}(\mathbf{R}_{2},\mathbf{r})$ due to a spin at
  $\mathbf{R}_{2}$ ($x=0$, $y=70~\mathrm{\AA }$).Same conventions and
  symbols as in Fig~\ref{fig:Two}.Compare to Fig.~\ref
  {fig:FourCrossR1}. }
\label{fig:FourCrossR2}
\end{figure}

The results of minimizing Eq. (\ref{eq:fourspins}) are shown in Fig.
\ref{fig:FourCrossR1} and \ref{fig:FourCrossR2}. Figure
\ref{fig:FourCrossR1} shows the spin density generated by a
perturbation at ${\bf R}_1$ and Figure \ref{fig:FourCrossR2} the spin
density when the magnetic perturbation is at ${\bf R}_2$.  
Note in Fig.~\ref{fig:FourCrossR1} that a
magnetic mirage appears at $\mathbf{R}_{3}$ ($x=-70~\mathrm{\AA }$,
$y=0$) but neither at $\mathbf{R}_{2}$ ($x=0,y=70~\mathrm{\AA }$) nor
at $\mathbf{R}_{4}$ ($x=0$, $y=-70~\mathrm{\AA }$).
Conversely in Fig. \ref{fig:FourCrossR2},
the magnetic mirage forms at $\mathbf{R}_{4}$ ($x=0$, $y=-70~\mathrm{%
  \AA }$) but neither at $\mathbf{R}_{1}$ ($x=70~\mathrm{\AA }$,
$y=0$) nor at $\mathbf{R}_{3}$ ($x=0$, $y=-70~\mathrm{\AA }$).
Therefore, the spin Hamiltonian formed by this structure has two pairs
of spins interacting independently. Each pair of spins do not couple
with the other pair, even though the interaction is mediated by an
electron gas that is shared by both pairs.  Such couplings have been
recently achieved experimentally\cite{eigler:03}, here we present a
structure that achieves a similar objective where the mirages are
much closer and couplings between crossing pairs are specifically avoided.

\subsection{Optimization for other purposes: optimal corrals to determine phase shifts}

Let us discuss for the moment the limitations of our approach. Our
model describes the resonances in the densities of states in terms of
an scattering theory which depends on a single parameter: the phase
shift $\delta$. The perturbation in the density of states due to single
impurity \cite{schneider:02} or an entire corral \cite{heller:464} has
been used to measure the characteristic phase-shift of a particular
type of impurity.  We note, however, that such a procedure is not very
sensitive to the phase shift and that might be the reason of some
controversy in the literature\cite{heller:464,harbury:96}.  The origin
of this lack of sensitivity comes from the fact that the systems
fitted were not optimized to be sensitive to the phase shift.  For
example, dense and closed shaped corrals do not produce 
densities of states that is sensitive to the phase shift of the
individual atoms of the corral.  It is therefore our goal in this
section to demonstrate the possibility of optimizing a quantum corral
for the determination of the phase shift $\delta$.

We first describe the optimization of a quantum corral
geometry with the aim of measuring $\delta$. Subsequently, we
characterize the corral as an experimental device to measure phase
shifts.

Let us first concentrate on Figure \ref{fig:ThreeBlockR1} that shows a
corral optimized to obtain a couple of mirages at symmetric points ${\bf
 R}_2$ and ${\bf R}_3$ due to a perturbation at ${\bf R}_1$. As we
noted in Section II Eq. (8), within linear response, magnetic mirages are
proportional to finite differences in the density of states under a
paramagnetic perturbation [i.e. $C_{\epsilon_{F}}({\bf R}_2,{\bf R}_1)\sim
\delta {\mathcal D}({\bf R}_2,{\bf R}_1)$] The latter have been measured
experimentally.\cite{manoharan:00,durkan:458}

As the corral in Figure \ref{fig:ThreeBlockR1} was forced to have
reflection symmetry along the $x$ axis, $\delta {\mathcal D}({\bf R}_2,{\bf
  R}_1)$ and $\delta {\mathcal D}({\bf R}_3,{\bf R}_1)$ are identical.
However, if we place an additional impurity outside the $x$ axis at a
point ${\bf R}_4$, the symmetry of the system is broken. A difference
in the amplitudes of the mirages at ${\bf R}_2$ and ${\bf R}_3$ due to
a perturbation at ${\bf R}_1$ will reflect the contribution of this
single asymmetric impurity at ${\bf R}_4$. Intuitively one would guess
that the difference will be largest if we put {\bf $R_4$} at $({\bf
  R}_1+{\bf R}_2)/2$ in order to interfere with the coupling between
${\bf R}_1$ and ${\bf R}_2$ but not with the coupling between ${\bf
  R}_1$ and ${\bf R}_3$. However, in order to help intuition, we
optimized the corral further so as it gives a large difference
$\Delta^2= \delta {\mathcal D}({\bf R}_2,{\bf R}_1)- \delta {\mathcal D}({\bf
  R}_3,{\bf R}_1)$ when there is a fix impurity at ${\bf R}_4$ while
the rest of the impurities have reflection symmetry along the $x$ axis
and we perturbed the system with an impurity at ${\bf R}_1$.  With
that goal we design the following cost function:
\begin{eqnarray}
\label{eq:costdiff}
E_4 &= &          
        -C_{\epsilon _{F}}({\bf R}_2,{\bf R}_1)
        -C_{\epsilon _{F}}({\bf R}_3,{\bf R}_1)
        +\left|
              C_{\epsilon _{F}}({\bf R}_2,{\bf R}_3)
              \right| \nonumber \\
    & &    -\left|
              ( C_{\epsilon _{F}}({\bf R}_2,{\bf R}_1)+C_{\epsilon _{F}}({\bf R}_2,{\bf R}_4)) \right.\nonumber \\
    & & \left.
             -(C_{\epsilon _{F}}({\bf R}_3,{\bf R}_1)+C_{\epsilon _{F}}({\bf R}_3,{\bf R}_4))
             \right| \times 0.5. 
\end{eqnarray} 
The first part of Eq. (\ref{eq:costdiff}) is the same as Eq.
(\ref{eq:threespins}). The second part is maximum when there is a big
difference in the mirages at ${\bf R}_1$ and ${\bf R}_3$ when there is an
additional impurity at ${\bf R}_4$ (treated in the limit of linear
response). The weight factor (0.5) in the last term in Eq.
(\ref{eq:costdiff}) controls the relative importance of large mirages
vs. large differences.

\begin{figure}
\includegraphics[width=0.95\linewidth,clip=true]{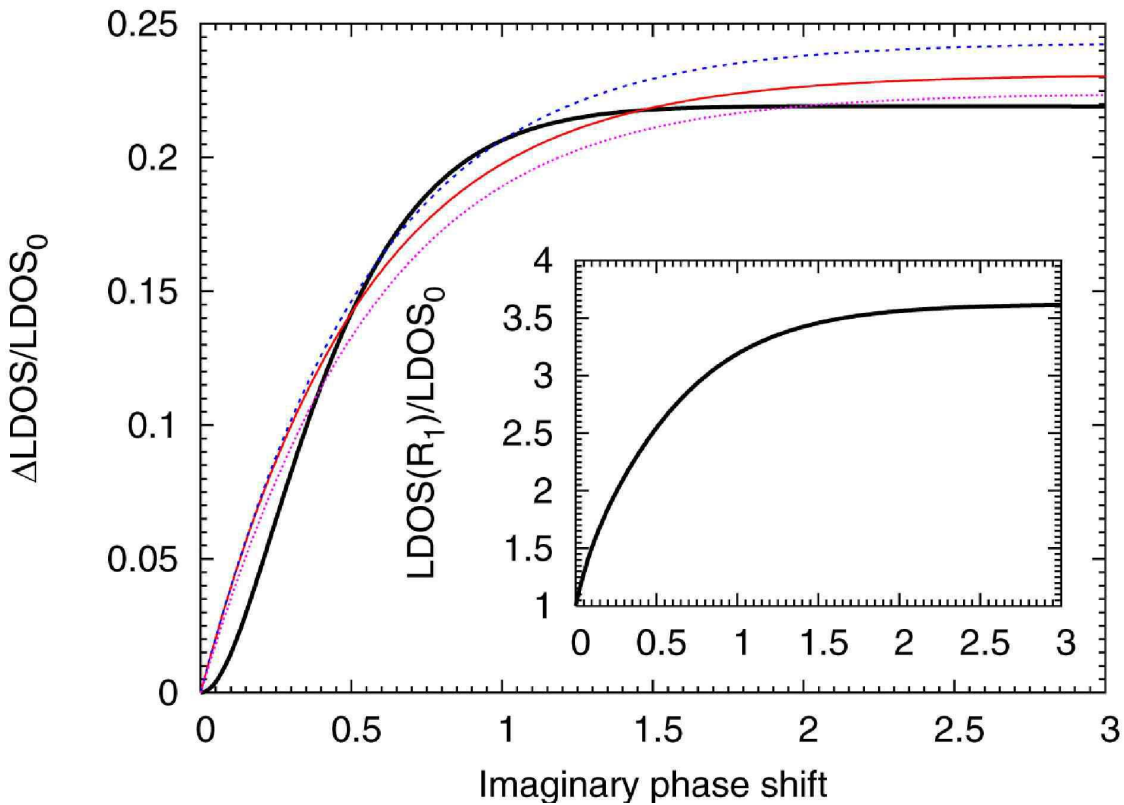}
\caption{\label{fig:MeasurePS} (Color online) 
$\Delta^2=\delta {\mathcal D}({\bf R}_2)-\delta {\mathcal D}({\bf R}_3)$ difference produced by an
additional impurity at ${\bf R}_4$ as a function of the phase shift of
all impurities (full line), or assuming a fixed value of the phase
shift of the corral (red $\theta= 0.5 i$, blue $\theta=1 i$, magenta
$\theta=2 i $ ) but varying the phase shift of the impurity at ${\bf
R}_4$. Units in density of state of free electron gas.}
\end{figure}

After the simulated annealing procedure, the corral obtained with the
cost function (\ref{eq:costdiff}) is similar to the one obtained in
Fig.  \ref{fig:ThreeBlockR1} because the cost function
(\ref{eq:threespins}) and (\ref{eq:costdiff}) are similar \cite{coordinates}.
  Let us concentrate on the amplitudes
of the mirages ($\delta {\mathcal D}$) as
a function of the phase shift $\delta$ of the impurities of the
corral.  In the inset of Fig.  \ref{fig:MeasurePS}(a) we show the
amplitude of the mirage (either at ${\bf R}_2$ or ${\bf R}_3$) due to
a perturbation at ${\bf R}_1$ for the symmetric case (without the
impurity at ${\bf R}_4$) as a function of the phase shift parameter
$\delta$. $\delta {\mathcal D}({\bf R}_2,{\bf R}_1)$ is measured in units of
the density of states at the Fermi level of the unperturbed electron
gas.  We see that the amplitude of the mirages grows monotonically with
$\delta$ saturating around $2.5$~.  While the magnitude of $\delta
{\mathcal D}$ could be used to determine, by comparison with an experiment,
the parameter $\delta$, that would be not enough to verify the
validity of the model used to construct the corral. As for a different
model (i.e. a real phase shift) we would find a different way to fit the
same data: we need at least two measurements to verify the validity of
a model with a single free parameter.  In Figure
\ref{fig:MeasurePS} (black line) we plot $\Delta^2$ when there is
an additional impurity at ${\bf R}_4$ as a function of a common phase
shift $\delta$.  The value of $\Delta^2$ also increases monotonically
with $\delta$ but saturates earlier (around $\delta = 1$). Note that
the difference between the two mirages can be of the order of 20 \% of the
density of states at the Fermi level.  Therefore, we believe that the
contribution of a single impurity can be measured within experimental
resolution, providing alternative data to corroborate the present
surface scattering model.

Finally, let us characterize this double mirage structure as an
experimental device to measure surface scattering cross sections or
phase shifts of arbitrary impurities. Let us assume that the surface
of Cu is doped mainly with type one atom (lets say Co) but also that
there are additional minority impurities on the surface available.
Let us say that we are not even sure what is the phase shift of our
majority impurity. The following question arises: is this quantum
corral made of impurities of unknown phase shift good enough to
estimate the phase shifts of any other impurity?.  The doted lines
lines in Figure \ref{fig:MeasurePS} have been drawn assuming that the
impurities of the corral have a phase shift $\delta$ of 0.5 1 and 2 as
a function of the phase shift $\delta \prime$ of a {\em different} the
impurity in ${\bf R}_4$ and a perturbation with phase shift $\delta$
at ${\bf R}_1$.  We see that $\Delta^2$ is quite insensitive to the
phase shift of the impurities of the corral, suggesting that only the
geometry chosen for confinement determines $\Delta^2$. Therefore, a
single quantum corral device can be used to ``measure'' the phase
shift of every impurity in the periodic table in spite of the
uncertainly in the knowledge of the phase shift of the impurities in
the corral.  Note finally, that the cost function was build such that 
the corral was very sensitive to small perturbations at $R_4$.
Therefore, our proposed ``experimental device'' is very sensitive to
small phase shifts and saturates faster for larger ones than the
amplitude of the mirages (see inset).  Alternatively one might have
chosen a different cost function to selectively measure higher vales
of $\delta \prime$.

\section{Discussion and Conclusions}

We have demonstrated the possibility to design theoretically quantum
corrals with multiple quantum mirages. In this work we have overcome
the problem posed by resonances and multiple scattering in quantum
corrals that give rise to a both a very large configuration space and
a very irregular response function.  This response function structure
makes deterministic algorithms fail to find good minima.  The
automated design algorithm is useful to attack the inverse problem
specially in cases where the intuition fails or gives poor answers.
Instead of covering the huge space of all possible configurations, we
propose a non deterministic method to explore the possible
configurations.  For the examples covered here, $10^6$ steps were
enough to optimize the structure.

The enhancement of magnetic interactions due to electronic
confinement together with the automatic design of the quantum corrals
allows us to tailor structures that generate a variety of spin
Hamiltonians involving several magnetic neighbors.  We demonstrate in
this work that it is in principle possible to design quantum corrals
such as the geometry of the system produces magnetic images at
predetermined points. This opens the possibility to actually construct
experimentally magnetic Hamiltonians of any sort imposed a priori.
This could open a path to study experimentally a number of
Hamiltonians which have been subject of intense theoretical research
along the years. In particular one could study one dimensional spin
Hamiltonians where there are analytical solutions.  Such ability to
construct Hamiltonians might be relevant in the context of
quantum computation.  We note in passing that one of the proposals to
build quantum computers is based on the interaction of nuclear spins
mediated by an electron gas\cite{kane:98}. The physical realization of
prescribed Hamiltonians is a necessary step in quantum computation.
The measurement of isolated spins in quantum corrals has been
experimentally demonstrated.~\cite{manoharan:00,durkan:458} The
exchange Heisenberg coupling was theoretically shown to be suitable
for quantum computation\cite{divincenzo:339}.  The combination of
these two ingredients and our method to design optimum quantum corrals
opens the possibility to actual realizations of quantum computers with
spins on a surface.  But, in addition, Quantum Computation requires
the ability to turn interactions on and off~\cite{divincenzo:339}. In
principle the interactions between impurities can be turned on and off
by changing the position of the Fermi level in the surface band with
respect to the bulk band. While
the control of the Fermi level might be difficult in copper surfaces
it is much easier in semiconductor quantum wells.

\section{Acknowledgments}
The authors would like to thank Giulia Galli for support comments and
suggestions and to Roy Pollock for a critical reading of the
manuscript. This work was performed under the auspices of CONICET
Argentina and of the U.S. Dept. of Energy at the University of
California/Lawrence Livermore National Laboratory under contract no.
W-7405-Eng-48.

\newpage 

\end{document}